\documentclass[prl,manuscript,amssymb,amsfonts,superscriptaddress]{revtex4}
\usepackage{epsfig}
\usepackage{amsmath}

\begin{document}

\title{Nonlinear dynamics of a dense two-dimensional dipolar exciton gas}

\author{Ronen Rapaport, Gang Chen, and Steven H. Simon}
 \affiliation{Bell Laboratories, Lucent Technologies,
600 Mountain Avenue, Murray Hill, New Jersey 07974}

\begin{abstract}

We use a simple model to describe the nonlinear dynamics of a
dense two dimensional dipolar exciton gas. The model predicts an
initial fast expansion due to dipole-dipole pressure, followed by
a much slower diffusion. The model is in very good agreement with
recent experimental results. We show that the dipole pressure
induced expansion strongly constrains the time available for
achieving and observing Bose-Einstein quantum statistical effects,
indicating a need for spatial exciton traps. We also suggest that
nonlinear ballistic exciton transport due to the strong internal
dipole pressure is readily achievable.
\end{abstract}

\maketitle

A new species of exciton, the spatially indirect dipolar exciton,
has recently emerged in the continuing search for degenerate
quantum gases and liquids in excitonic systems
\cite{SnokeScience2002,ButovJPCM2004}. Spatially indirect dipolar
excitons are coulomb bound electron-hole pairs in which the
electron and the hole are confined in two spatially separated
two-dimensional (2D) layers. Such excitonic states are usually
achieved in double quantum well (DQW) structures with an external
electrical bias applied perpendicular to the quantum well (QW)
plane to lift the degeneracy of the confined states in the two
respective QW's\cite{SnokeScience2002,ButovJPCM2004}. As a result,
the lowest energy optical transitions occur between the two QW's,
leading to the formation of indirect excitons. While the binding
energy of such an exciton is only slightly modified due to the
spatial separation, the lifetime is enhanced by many orders of
magnitude compared to direct excitons (where the electron and the
hole are in the same layer). This extremely long lifetime makes
them especially interesting in the context of degenerate quantum
gases of excitons, since it may provide sufficient time for
thermalization as well as cooling of initially hot excitons.
Another prominent feature of this type of indirect exciton is its
permanent dipole moment, aligned perpendicular to the QW plane,
due to the spatial separation of the electron and the hole. This
dipole moment gives rise to a repulsive dipole-dipole force
between excitons. On one hand, this repulsion prevents the
formation of e-h complexes such as biexcitons and e-h droplets,
and therefore makes such systems even more attractive candidates
for observing degenerate Bose-Einstein gases. On the other hand,
it leads to a strong driven expansion of a dense dipolar exciton
gas, as was very recently observed\cite{VorosPRL2005}. Hence, this
dipole-dipole repulsion seems to constrain the time that such an
indirect exciton gas can remain dense after photoexcitation, an
obvious obstacle to the observation of quantum statistical effects
of excitons.

Here, we use a simple model to describe the dynamics of such
dipolar excitons, taking into account the dipole-dipole
interaction. Besides quantitatively describing current
experimental observations of the nonlinear expansion dynamics of
dipolar exciton gases, this model suggests that the expansion
caused by the dipole repulsion indeed strongly limits the time
that a high mobility exciton gas remains dense, explaining the
difficulties in the observations of exciton condensation. Some
sort of spatial confining (exciton traps) might then be necessary
for achieving a dense and cold dipolar exciton gas, a prerequisite
for excitonic Bose-Einstein condensation in such systems.

Two recent experimental observations of dipolar exciton dynamics
by Voros et al. (Ref.~\cite{VorosPRL2005}) are the basis of our
model assumptions: (a) The dipolar exciton expansion dynamics can
be divided into two regimes: immediately after the laser
excitation pulse, a very fast expansion is observed, in which the
excited exciton cloud expands from its initial small diameter
($\sim 30\mu m)$ by a factor of $\sim 6$, in a few tens of
nanoseconds. At later times, the expansion slows down
significantly, and can be very well described by exciton diffusion
process. (b) The diffusion coefficient of the dipolar excitons,
$D_X$, has a power law dependence on the well width, $L$, where it
is found to be of the form $D_X \propto L^6$. This specific
dependence was found previously for free electrons in similar
structures by Sakaki et al. in Ref.~\cite{SakakiAPL1987}, which is
well explained by interface roughness scattering from well width
fluctuations \cite{PrangePR1968,GoldSSC1986,GoldPRB1987}. This
finding is a strong evidence that for dipolar excitons at low
temperatures, potential fluctuations due to interface roughness is
also the dominant scattering mechanism.

Based on these two observations, we suggest a simple model to
describe the dynamics of the dipolar excitons (see also
Ref.~\cite{Ivanov}). The model includes the dipole-dipole
repulsion between the excitons as the main driving force for the
initial fast expansion, and the static interface roughness
scattering as the drag mechanism.

The dipole-dipole interaction energy per exciton of a gas of
perfectly aligned dipolar excitons is given by
\cite{Ivanov,BenTabouPRB2001}:
\begin{equation} \varepsilon_{dd}(\vec{r})=\frac{4\pi
e^2z_0}{\epsilon}n_X\equiv\alpha n_X, \label{ddenergy}
\end{equation}
where $n_X\equiv n_X(\vec{r})$ is the 2-dimensional density of the
dipolar excitons, $z_0$ is their dipole length, and $\epsilon$ is
the background dielectric constant. The resulting dipole-dipole
repulsion force,
\begin{equation}
\textbf{F}_{dd}=-\alpha\nabla n_X, \label{ddforce}
\end{equation}
proportional to the density gradient, will drive the exciton
expansion. The other effective force that is responsible for the
diffusion of an exciton gas is
\begin{equation}
\textbf{F}_D=-\nabla \zeta.
\end{equation}
Here $\zeta = kT_Xln\left(1-e^{-T_0/T_X}\right)$ is the chemical
potential in the non-interacting limit\cite{Ivanov}, where
$T_0=\frac{\pi \hbar^2n_X}{2km_X}$ is the degeneracy temperature
and $T_X$ and $m_X$ are the exciton gas temperature and exciton
effective mass, respectively. The overall internal force driving
the excitons is therefore
$\textbf{F}=\textbf{F}_{dd}+\textbf{F}_D$.
 The resulting exciton current
$\textbf{J}=n_X\mu \textbf{F}=n_X\mu
(\textbf{F}_{dd}+\textbf{F}_D)$ consists of contributions from
both the dipole-dipole repulsion and the diffusion:
\begin{eqnarray}
\textbf{J}_d&=&-n_X\mu \alpha \nabla n_X  \label{dd-current}\\
\textbf{J}_D&=&-n_X \mu \nabla
\zeta=-\mu\frac{kT_0}{e^{T_0/T_X}-1}\nabla n_X\label{diff-current}
\end{eqnarray}
Here, $\mu$ is the exciton mobility which is, in general, a
function of $\left|\textbf{F}\right|\equiv F$.

\begin{figure}[htb]
\begin{center}
\includegraphics[scale=0.9]{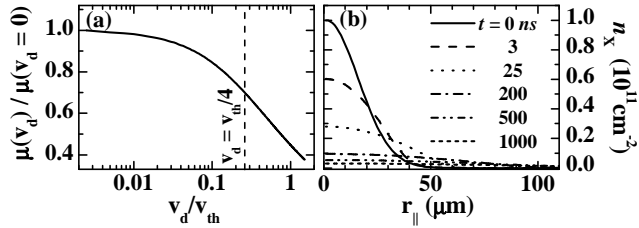}
\caption{(a) Normalized dipolar exciton mobility as a function of
the ratio of the drift velocity to the thermal velocity
$v_d/v_{th}$. (b) Dipolar exciton radial density distribution
profiles for different times.}\label{figure1}
\end{center}
\end{figure}

We now construct a simplified model to obtain the dependence of
$\mu$ on $F$ following the observation that interface roughness
scattering is the dominant relaxation mechanism. We model the
interface roughness scattering by a random distribution of
quasi-elastic, hard sphere scattering centers with a
characteristic mean free path $l$. In the low-field limit, the
mobility, $\mu_{lf}$, is independent of $F$. In that limit,
$\mu_{lf}$ is related to the diffusion coefficient through the
modified Einstein relation\cite{Ivanov} $D=\mu
kT_X/(e^{T_0/T_X}-1)$, which in the non-degenerate limit reduces
to the well-known $\mu_{lf} =D_X/kT_X$. In this picture,
$D_X=l^2/\tau=\it{l}v_{th}$, where $\tau$ is the exciton
scattering time and $v_{th}$ is the thermal velocity
$v_{th}=\sqrt{2kT_X/m_X}$. In the high field limit however, the
drift velocity $v_d$ depends on the force $F$ and is determined by
the average velocity the exciton gains between successive
collisions, given by $v_d=\sqrt{Fl/(2m_X)}$. The resulting high
field limit for the mobility is then
$\mu_{hf}=v_d/F=\sqrt{l/(2m_XF)}$. We thus construct a smooth
function for the mobility as a function of $F$ that conforms to
the above two limits:
\begin{equation}
\mu(F)=\sqrt{\frac{l/(2m_X)}{F+kT_X/4}}. \label{mobility}
\end{equation}
The crossover between the two transport regimes occurs at an $F$
which corresponds to $v_{th}\approx 4v_d$. The dependence of the
mobility on $v_d/v_{th}$ is plotted in Fig.~\ref{figure1}a. The
free exciton gas dynamics is now described by a nonlinear
diffusion-drift equation,
\begin{equation}
\frac{\partial n_X}{\partial t}+\nabla\cdot
(\textbf{J}_D+\textbf{J}_d)+\frac{n_X}{\tau_X}-I_X(r,t)=0,
\label{excitoneq}
\end{equation}
where $I_X(r,t)$ is the exciton source and $\tau_X$ is the exciton
recombination lifetime. This nonlinear equation can be solved
numerically.

Since $\mu$ depends on the internal force $F$, the exciton
\emph{transport} is in general nonlinear, analogous to nonlinear
I-V characteristics of charged particles. In the case that $v_d\ll
v_{th}$, $\mu$ becomes independent of $F$ and the exciton
transport becomes linear. However, even in the linear transport
limit, the exciton expansion can be very fast, strongly driven by
the dipole repulsion which can dominate over diffusion (it turns
out that this is the case in the experiments presented in
Ref.~\cite{VorosPRL2005}). Thus, we define the following quantity
to describe the importance of the fast dipole driven expansion:
\begin{equation}
\gamma(r,t)=J_d/J_D
=\frac{n_X\alpha\left(e^{T_0/T_X}-1\right)}{kT_0} \label{Jratio},
\end{equation}
which, in the dilute limit, reduces to just the ratio between the
dipole-dipole energy term and the thermal energy of the excitons,
$\gamma=\frac{\alpha n_X}{kT_X}$. When $\gamma\gg 1$, the
expansion is driven by the repulsive dipole-dipole force. When
$\gamma\ll 1$, the expansion becomes diffusive. An exciton gas
will exhibit a fast driven expansion if its initial density
profile, $n_X^0\equiv n_X(r=0,t=0)$, satisfies
$\gamma(r=0,t=0)>1$. As a dense exciton gas starts to rapidly
expand due the dipole forces, its density will decrease and the
expansion will slow down. Wherever the density decreases such that
$\gamma(r=0,t=0)<1$, the expansion will become diffusive, until
eventually the diffusion will dominate the dynamics of the whole
exciton gas.

We start our discussions by comparing the model calculations to
the experimental results of Ref.~\cite{VorosPRL2005}. There, DQW
structures were used with different well widths: 80A, 100A, 120A,
and 140A and the corresponding $D_X$ were found to be (in units of
$cm^2/s$): 0.24, 0.74, 2.08, and 9.4. These numbers yield a
corresponding mean free path of $l(nm)=1.4,4.5,12.5,56.6$, for the
experimental temperature, $T=2K$, and for an exciton effective
mass $m_X=0.2m_e$. Note that for the wide DQW structures $l>a_X$
(where $a_X\approx 14nm$ is the exciton Bohr radius) while for the
narrow DQW's, $l<a_X$. Using the parameters of
Ref.~\cite{VorosPRL2005}, we find that exciton drift velocity
$v_d$ is always smaller than $v_{th}$ even for samples with the
highest $D_X$. The exciton transport is therefore linear with
$\mu\simeq \mu_{lf}$. However, we estimate that
$\gamma(r=0,t=0)\simeq 800$, indicating that the exciton cloud
should exhibits an initial fast driven expansion due to the dipole
pressure, and later a diffusive behavior.

For our calculations, we assume radial symmetry and an initial
gaussian distribution of optically excited dipolar excitons, with
a half width $\sigma_0 =15\mu m$ and an $n_X^0=10^{11} cm^{-2}$
(unless stated otherwise), similar to the experimental conditions
of Ref.~\cite{VorosPRL2005}. We also use an exciton dipole length
of $z_0=12nm$. A temperature of 3K and an exciton lifetime of
$3\mu s$ are assumed. Fig.~\ref{figure1}b presents calculated
exciton density profiles for several different times and for
$D_X=9.4 cm^2/s$, showing the exciton expansion.

\begin{figure}[htb]
\begin{center}
\includegraphics[scale=0.3]{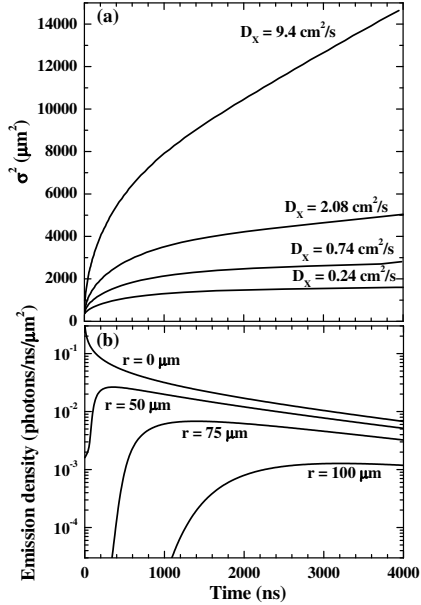}
\caption{(a) $\sigma^2$ of the density distribution profiles as a
function of time, for the diffusion coefficients of
Ref.~\cite{VorosPRL2005}. (b) Calculated dipolar exciton emission
density as a function of time from different radial positions of
the exciton cloud, corresponding to the experimental positions
measured in Ref.~\cite{VorosPRL2005}. The parameters for the
calculations in both (a) and (b) are given in the
text.}\label{figure2}
\end{center}
\end{figure}

Fig.~\ref{figure2}a plots the calculated squares of the half-width
- half-maximum ($\sigma^2$) of the exciton density profiles at
various times, for various $D_{X}$ extracted from the experiments
of Ref.~\cite{VorosPRL2005}. Both the fast initial expansion and
the later slow diffusive regime are clearly seen. The larger the
$D_{X}$, the faster the initial expansion, as $\mu$ and $D_{X}$
are linked through the scattering length. While we did not try to
do any "fine tuning" of parameter fitting, there is a very good
agreement with the experimental results (Fig. 4 in
Ref.~\cite{VorosPRL2005}). The only parameters for the calculation
that were not predetermined by the experiments are the exact
initial exciton density and the exact exciton gas temperature. The
density can be only roughly estimated from the excitation
intensities quoted in Ref.~\cite{VorosPRL2005}, which indeed
yields $n_X^0\sim 1\times 10^{11} cm^{-2}$.

\vspace*{0cm}

The experimental data, however, seem to have a sharper transition
from the driven regime to the diffusive one than the calculation.
This is seen in the sharper bend of $\sigma^2$ with time shown in
Ref.~\cite{VorosPRL2005}. This discrepancy may be due to the
heating of the exciton gas as it expands, transferring the
potential energy into thermal energy by the elastic scattering
processes. Such heating process is not included in our model.

Fig. \ref{figure2}b presents the calculated time traces of the
emission from different fixed radial positions of the expanding
exciton gas (for $D_X=0.74 cm^2/s$). These calculated time traces
are also in a good agreement with the experiments (Fig. 2 in
Ref.~\cite{VorosPRL2005}). It seems then that this model can quite
accurately approximate real experimental observations, based on
the few simple assumptions given above.

\vspace*{0cm}

\begin{figure}[htb]
\begin{center}
\includegraphics[scale=0.35]{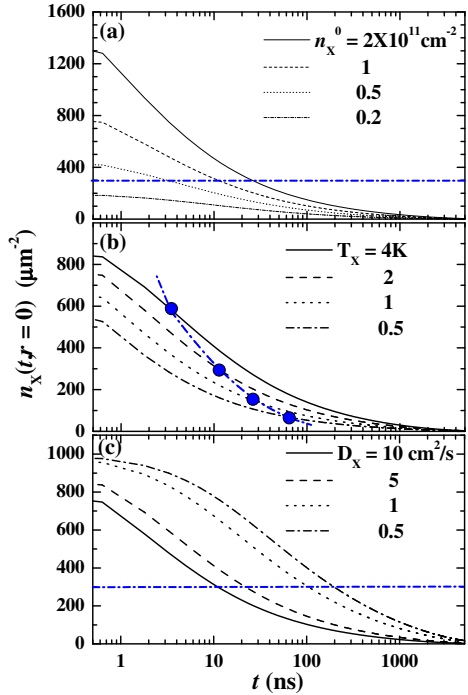}
\caption{Center exciton densities as a function of time,
$n_X(t,r=0)$ for (a) different initial center densities $n_X^0$,
(b) different exciton gas temperatures $T_X$, and (c) Different
exciton diffusion coefficients $D_X$. The relevant parameters for
the calculations in (a), (b) and (c) are given in the text. The
dashed-dot lines in (a) and (c) and the circles in (b) mark the
de-Broglie density $n_X^c$.}\label{figure3}
\end{center}
\end{figure}

After establishing the validity of this picture, we are now ready
to discuss some of its important consequences. It is already clear
that a dense and mobile dipolar gas will tend to expand quickly
due to the internal repulsive forces. If the initial cloud is
small, the expansion will quickly reduce the exciton gas density
and will limit the time the high density condition can be
sustained. We start by discussing small initial exciton clouds,
excited by tightly focused laser beams, which is the most common
experimental approach. In Fig. \ref{figure3}a we plot the central
density of an expanding, initially small ($\sigma=15\mu m$), high
mobility exciton gas ($D_X=10cm^2/s$). The calculations are
performed with a gas temperature of $T_X=2K$, as a function of
time, for different initial exciton densities $n_X^0$. the
horizontal dash-dotted line represents the density, $n_X^c$, in
which the exciton thermal de-Broglie wavelength is equal to the
exciton inter-particle distance, i.e.,
$\lambda_{dB}^2n_X^c/g_X=1$. One expects to observe quantum
statistical effects due to Bose-Einstein distribution only for
densities larger than $n_X^c$. As can be seen, within less than
$30ns$, the density drops below $n_X^c$ even for high initial gas
densities. This time is much shorter than the exciton lifetime of
$3\mu s$. Again, this is a consequence of the fast initial
expansion. Fig. \ref{figure3}b shows the similar density-time
traces for an initial density $n_X^0=10^{11}cm^{-2}$ for different
temperatures. The circles on each density-time trace mark $n_X^c$
for the particular temperature, given by:
$n_X^c=m_XgkT_X/(2\pi\hbar^2)$, and the curved dashed-dot line is
a guide to the eye. While decreasing temperature results in an
increased mobility for a given $D_X$, there is also a linear
decrease in $n_X^c$, resulting in an improvement of the time
interval in which $n_X>n_X^c$. However, even at $T_X=0.5K$ this
time is still under $70ns$, significantly shorter than the exciton
radiative lifetime, $\tau_X$. Fig. \ref{figure3}c shows similar
traces for different diffusion coefficients. A smaller diffusion
coefficient is related to a smaller mean free path $l$, which in
turn yields less expansion and less density decrease.

\vspace*{1cm}
\begin{figure}[htb]
\begin{center}
\includegraphics[scale=0.85]{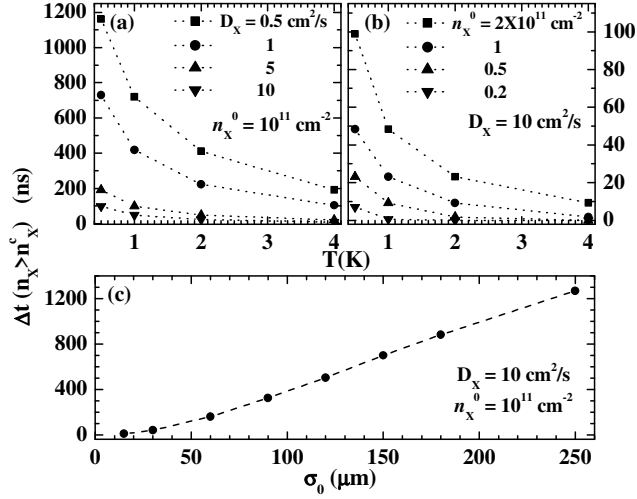}
\caption{Time interval $\Delta t$ in which the center dipolar
exciton gas density, $n_X(r=0)$ is larger than the de-Broglie
density, $n_X^c$, as a function of the exciton gas temperature,
$T_X$. (a) shows the results for different diffusion coefficients,
$D_X$ while (b) shows different initial center densities, $n_X^0$.
(c) shows $\Delta t$ as a function of the initial size of the
exciton cloud for $T_X=2K$.}\label{figure4}
\end{center}
\end{figure}

Fig. \ref{figure4}a,b summarize the above results. The time
interval $\Delta t$ in which $n_X>n_X^c$ as a function of the
exciton gas temperature, for various $D_X$ values is shown in Fig.
\ref{figure4}a and for various initial densities $n_X^0$ in Fig.
\ref{figure4}b. For high mobility (diffusivity) samples,
decreasing the temperature and increasing the initial density only
slightly increases $\Delta t$, to values still much smaller than
$\tau_X$. This is due to the fast driven expansion. For low
mobility samples, at very low temperatures, $\Delta t$ can be in
the microsecond range, comparable to $\tau_X$. For these samples,
however, the mean free path, $l$, is much smaller than the exciton
Bohr radius, $a_X$. This may complicate the observation of
Bose-Einstein quantum statistical effects in general and excitonic
BEC in particular due to the disorder induced localization of the
excitons.

Fig. \ref{figure4}c shows the dependence of $\Delta t$ on the
initial size of the exciton cloud or $D_X=10 cm^2/s$ and $T_X=2K$.
As the initial size increases (keeping the same center density),
the dipole driving force decreases $F_{dd}\propto |\nabla
n_X|\propto 1/\sigma_0$, and so does the expansion rate. Starting
with large enough initial exciton clouds leads to a much smaller
expansion rates and thus to $\Delta t$ values comparable to the
exciton lifetime $\tau_X$.

To conclude this analysis, there are two strategies to keep  a
high mobility dipolar exciton gas dense enough for long enough
time intervals. The first is to optically excite an initially very
large cloud, thus suppressing the fast expansion. This approach,
however, is not favorable since it involves high excitation power
which could lead to complications such as increased heating of the
sample. The other strategy is to design an artificial spatial
confinement in the QW plane. For such dipolar excitons, the
trapping potential can be induced either through a local strain in
the QW \cite{SnokeSSC2005} or by locally applying the external
electrostatic potential\cite{RapaportPRB2005}. These exciton traps
not only keep the excitons dense, but also eliminate the extra
heating (due to the dipole-dipole potential) of a free-expanding
exciton gas mentioned previously and breaking the 2D translational
which allows for the the formation of BEC.

It is interesting to compare the two characteristic densities of
this problem. The first is the de-Broglie density, $n_X^c$, that
marks the onset of a distinguishable Bose-Einstein statistics
effects. The second density is the critical density for the driven
expansion regime, $n_d^c=kT_X/\alpha$, derived by setting
$\gamma=1$ in the linear transport limit and assuming
non-degenerate gas. $n_d^c$ marks the transition from a fast
dipole-pressure driven exciton expansion to a diffusive expansion.
Their ratio yields,
\begin{equation}
\frac{n_X^c}{n_d^c}=\frac{2m_Xge^2z_0}{\hbar\epsilon}\approx 20,
\end{equation}
for typical GaAs DQW structures and is independent of temperature,
justifying the above assumption. This means that a
\emph{degenerate} Bose-Einstein gas of dipolar excitons is
\textit{always} strongly driven by the dipolar pressure. This can
have some interesting consequences. For example, it indicates that
a highly degenerate \emph{dipolar} gas will expand quickly, in
contrast to most atomic BEC gases that tend to expand very slowly.
Also, it is not clear how a driven expansion of a highly
degenerate exciton gas would differ from that of a classical gas
and how would the scattering of such an exciton gas from well
width fluctuations would differ from the simplistic, ballistic
"billiard ball" scattering model we have used. Our model also
suggests that for high mobility samples, such as the ones used in
Ref.~\cite{VorosPRL2005}, the mobility is on the edge of the
nonlinear transport regime. By cooling the exciton gas further or
by growing wider samples the predicted range for the study of
highly nonlinear transport due to the internal dipole pressure can
be reached for the first time.

We thank Loren Pfeiffer, P.M. Platzman, David Snoke and Xing Wei
for enlightening discussions.


\begin{thebibliography}{11}
\bibitem{SnokeScience2002} D. Snoke, Science \textbf{298}, 1368 (2002).
\bibitem{ButovJPCM2004} L.~V. Butov, J. Phys.: Condens. Matter.
\textbf{16} R1577 (2004).
\bibitem{VorosPRL2005} Z. V\"{o}r\"{o}s, R. Balili, D.W. Snoke, L.~N.
Pfeiffer, and K. West, Phys. Rev. Lett. \textbf{94}, 226401
(2005), cond-mat/0504151 (2005).
\bibitem{SakakiAPL1987} H. Sakaki, T. Noda, K. Hirakawa, M.
Tanaka,  and T. Matsusue, Appl. Phys. Lett. \textbf{51}, 1934
(1987).
\bibitem{PrangePR1968} R.E. Prange and T. W. Nee, Phys. Rev.
\textbf{168}, 779 (1968).
\bibitem{GoldSSC1986} A. Gold, Solid State Commun. \textbf{60}, 531
(1986).
\bibitem{GoldPRB1987} A. Gold, Phy. Rev. B \textbf{35}, 723 (1987).
\bibitem{Ivanov} A. Ivanov, P.B. Littlewood and H. Haug, Phys. Rev. B \textbf{59}, 5032 (1999); A. Ivanov, Europhys. Lett. \textbf{59}(4), 586
(2002).
\bibitem{BenTabouPRB2001} S. Ben-Tabou de-Leon and B. Laikhtman,
Phys. Rev. B \textbf{63}, 125306 (2001).
\bibitem{SnokeSSC2005} D. Snoke, Y. Liu, Z. Voros, L. N. Pfeiffer,
and K. West, Solid State Commun. \textbf{134}, 37 (2005).
\bibitem{RapaportPRB2005} R. Rapaport, G. Chen, S. Simon, O. Mitrofanov, L. N. Pfeiffer, and P. M.
Platzman, Phys. Rev. B, \textbf{72}, 000000 (2005),
Cond-mat/0504178, (2005).
\end{thebibliography}
\end{document}